\documentclass[12pt]{article}
\usepackage[a4paper, margin=2.5cm]{geometry}
\usepackage{amsmath, amssymb}
\usepackage{graphicx}
\usepackage{hyperref}
\usepackage{color}

\title{An Overview of Classical Scattering by a Lennard-Jones Potential}
\author{Tiago Kroetz\\ \small Universidade Tecnológica Federal do Paraná, Pato Branco, PR, Brasil\\ \small \texttt{kroetz@utfpr.edu.br}}
\date{}

\begin{document}

\maketitle
\begin{center}
\small
\textbf{Published as:} Tiago Kroetz, \textit{Eur. J. Phys.} \textbf{46}, 055001 (2025).\\
DOI: \href{https://doi.org/10.1088/1361-6404/adecce}{10.1088/1361-6404/adecce}
\end{center}
\vspace{0.5cm}

\begin{abstract}
We present a numerical and conceptual overview of classical scattering under a Lennard-Jones potential. All results were obtained by direct integration of particle trajectories in the two-dimensional plane for varying impact parameters and energies. From these trajectories, we compute deflection functions, identify critical values associated with unstable circular orbits, and visualize non-monotonic behavior in the angular deflection. We construct color maps representing the deflection angle and interaction delay time as functions of energy and impact parameter, and determine the differential cross section from the deflection function, accounting for multi-valued contributions. These results are displayed both as individual plots and as a full color map in the energy–angle plane, providing a detailed picture of the scattering behavior in this central potential.
\end{abstract}

\textbf{Keywords:} Classical scattering, Lennard-Jones potential, differential cross section, effective potential

\section{Introduction}

Classical scattering theory has played a central role in the development of physics, from Rutherford’s landmark experiments to modern molecular dynamics. Recent works, such as Žugec and Topić~\cite{zugec2020}, have revisited the geometrical aspects of repulsive Rutherford scattering from an educational perspective, reinforcing the didactic value and ongoing relevance of classical scattering analyses. The standard textbook treatment often focuses on repulsive Coulomb interactions, which, while analytically elegant, lack the physical richness found in realistic intermolecular potentials. One such potential is the Lennard-Jones (LJ) potential \cite{lennardjones1924}, widely used to model interactions between neutral, nonpolar atoms or molecules, such as noble gases. It is defined as:
\begin{equation}
V(r) = 4\varepsilon \left[ \left(\frac{\ell}{r}\right)^{12} - \left(\frac{\ell}{r}\right)^6 \right]
\label{potLJ}
\end{equation}

This potential captures both short-range repulsion, representing Pauli exclusion due to overlapping electron clouds, and long-range attraction, representing van der Waals interactions. Its importance is further emphasized by its historical and modern applications, including transport theory~\cite{hirschfelder1948}, molecular dynamics~\cite{stephan2019}, nanoparticle collisions~\cite{nietiadi2022}, and educational simulations~\cite{sperandeo1987}.

Despite its simplicity, the Lennard-Jones potential continues to be employed in computational implementations, such as DSMC methods~\cite{venkattraman2012}, and in the derivation of transport coefficients~\cite{khrapak2014b, laricchiuta2007, pirani2008}. These examples illustrate its robustness and relevance. In a related study, Khrapak~\cite{khrapak2014c} demonstrated that the deflection function for strongly attractive potentials—including Lennard-Jones—exhibits a quasi-universal shape when normalized by the critical impact parameter, enabling simplified scaling for transport integrals across different potentials.

The appeal of using the Lennard-Jones potential in a classical context lies in its ability to reveal rich trajectory behaviors—such as orbiting, retroreflection, and critical scattering with an intuitive geometrical framework. In this work, we revisit the classical scattering problem under a Lennard-Jones potential using direct numerical integration of the particle trajectory. Unlike treatments focused solely on deflection functions, we also present color maps of scattering angle and interaction delay time as functions of energy and impact parameter, revealing non-monotonicities and divergence regions. Differential cross sections are computed from the deflection function, taking into account multiple values of impact parameter contributing to the same scattering angle. Although the theoretical foundations of scattering (deflection angle, impact parameter, and cross section) are well established~\cite{goldstein2002, balint2015, Landau}, our presentation focuses on visual and intuitive representations to improve understanding. 

We justify the classical approach by noting that in the regime where the de Broglie wavelength is small compared to the interaction scale (i.e., $mv\ell \gg \hbar$), quantum effects are negligible. This inequality can also be expressed in terms of temperature as $T \gg \hbar^2 / (m k_B \ell^2)$. For argon, using $\ell = 3.405 \text{\AA}$ \cite{balint2015} and $m = 6.63 \times 10^{-26}\, \text{kg}$, we obtain $T \gg 0.16\, \text{K}$, confirming that the classical approximation is valid for essentially any realistic thermodynamic condition. Moreover, the classical framework provides geometrically intuitive interpretations of particle trajectories, deflection angles, and critical scattering behavior, which are less transparent in a quantum mechanical treatment.

The remainder of this article is organized as follows. In Section~2, we briefly review the foundations of classical scattering theory, including key concepts such as impact parameter, deflection angle, and differential cross section. Section~3 introduces the Lennard-Jones potential and discusses the corresponding effective potential, highlighting the role of angular momentum conservation. Section~4 presents numerical results for classical trajectories and the deflection function, emphasizing the transition between monotonic and non-monotonic regimes. In Section~5, we analyze the scattering process in greater depth through color maps of scattering angles and delay times, and compute the differential cross section from the deflection function. Finally, Section~6 summarizes the main findings and highlights the relevance of this study to physics education and to the conceptual understanding of molecular scattering processes involving Lennard-Jones interactions.

\section{Classical Scattering Theory}

The fundamental elements of classical scattering under central forces are well established~\cite{goldstein2002, balint2015, Landau}. In such systems, the force depends only on the radial distance $r$ from the center of the potential, and as a result of the conservation of the angular momentum, the motion of the scattered particle is confined to a plane. The analysis can therefore be restricted to two spatial dimensions.

\begin{figure}[h!]
\centering
\includegraphics[width=0.8\textwidth]{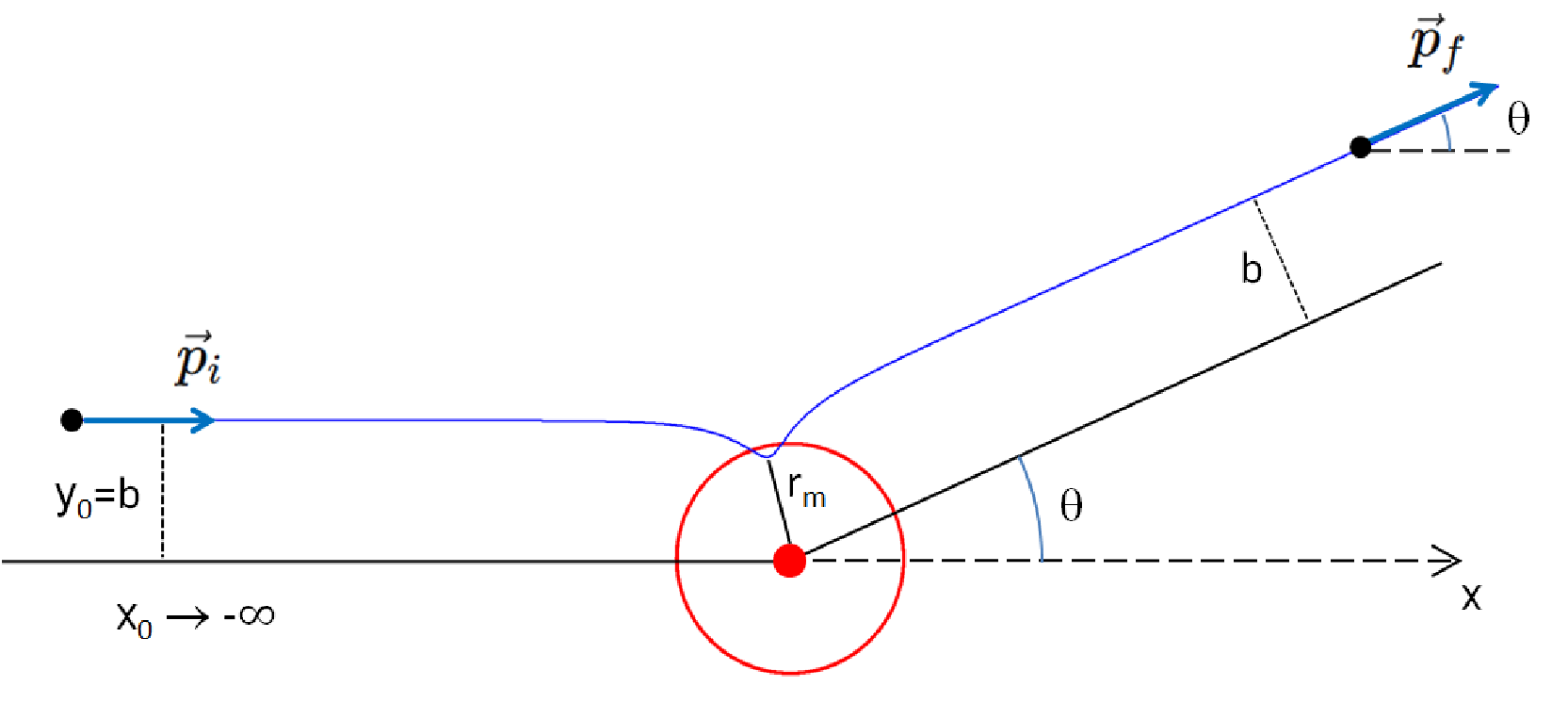}
\caption{Trajectory of a classical particle scattered by a central potential with both attractive and repulsive regions. The impact parameter $b$, the scattering angle $\theta$, and the periapsis distance $r_m$ are indicated. The red hollow circle marks the distance at which the interaction force changes sign, separating the attractive outer region from the repulsive inner region. The solid red disk represents the scattering center (not to scale)}.

\label{fig:scattering}
\end{figure}

Figure~\ref{fig:scattering} schematically illustrates the geometry of a scattering event. A particle approaches from infinity with initial momentum $\vec{p}_i$ and an impact parameter $b$, defined as the perpendicular distance from the initial trajectory to the center of the potential. The trajectory is initially a straight line, as the potential is negligible far from the interaction region. The central interaction then deflects the path, and the trajectory returns to a straight line after the particle escapes the effective range of the force. The angle between the initial and final directions of motion is the scattering angle $\theta$, which is the measurable deviation in the laboratory frame and, due to axial symmetry, is restricted to the range $[0, \pi]$. The change in direction results from the central force, but the interaction is conservative. In the fixed-target laboratory frame adopted here, the kinetic energy and the magnitude of the linear momentum of the scattered particle remain constant, with only the direction of motion being altered.

The interaction is fully characterized by the particle’s energy $E$ and its impact parameter $b$, leading to a well-defined scattering angle function $\theta(E,b)$. The point of closest approach between the particle and the target is referred to as the periapsis and is denoted $r_m$. The figure represents a generic attractive-repulsive central potential, such as the Lennard-Jones potential, where the red circle marks the approximate onset of the strongly repulsive core. For certain values of the impact parameter, the particle may undergo orbital-like revolutions around the scattering center before escaping, indicating a transient trapping in the attractive well. 

\begin{figure}[ht]
    \centering
    \includegraphics[width=0.6\linewidth]{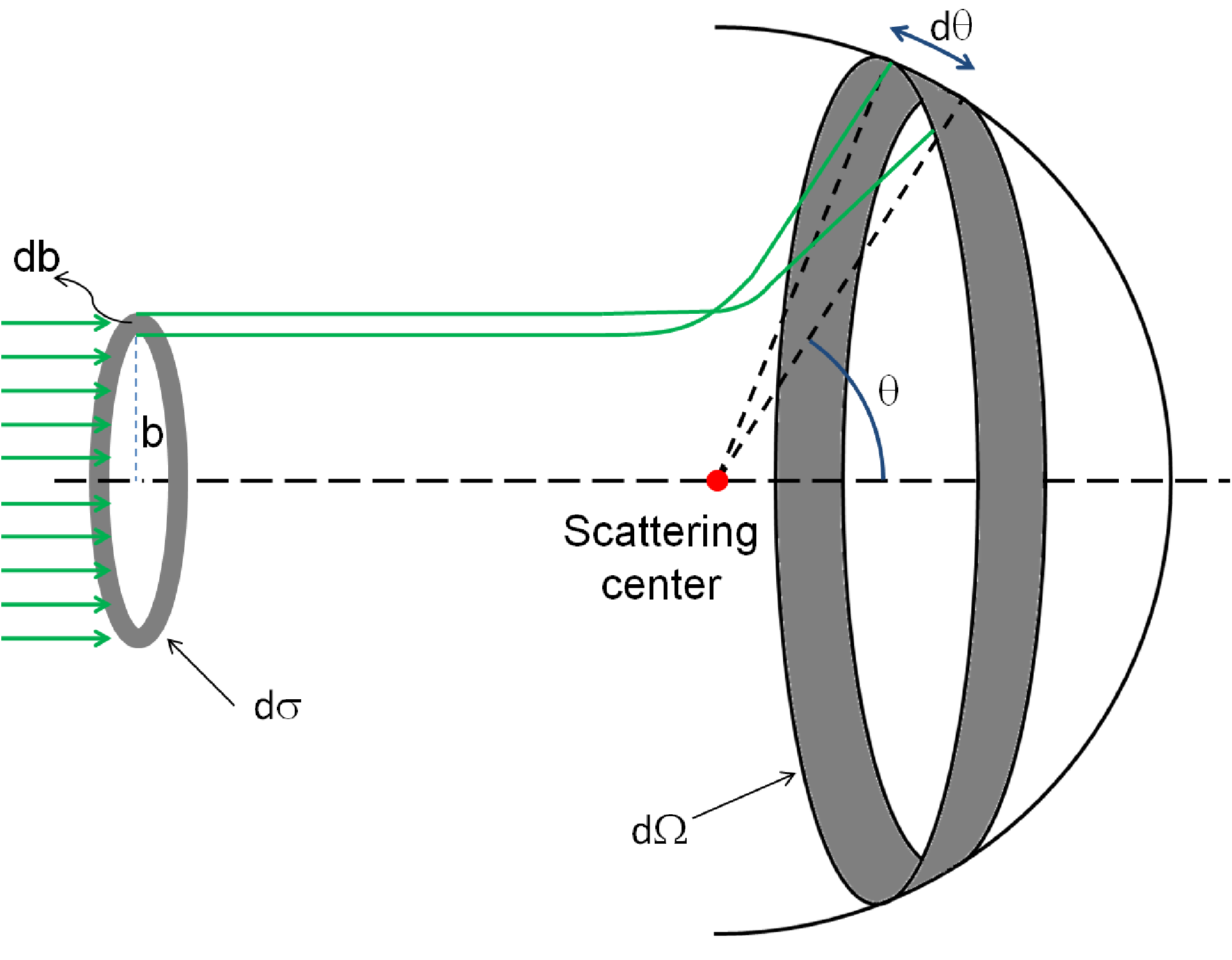}
    \caption{"Scattering from an annular element of the incident beam, with impact parameter $b$ and thickness $db$, into an axially symmetric solid angle element $d\Omega$, centered at scattering angle $\theta$.}
    \label{fig:solidangle}
\end{figure}

While the function $\theta(E,b)$ encapsulates the microscopic dynamics of individual trajectories, experimental observations rely on the intensity of scattered particles per angular sector, since neither the impact parameters nor the corresponding trajectories are accessible in practice. Therefore, a macroscopic description based on angular distributions is more appropriate. To describe the scattering in terms of experimentally measurable quantities, we define the differential cross section $d\sigma$ as the infinitesimal area of the incoming beam whose particles are scattered into a given solid angle element $d\Omega$. This is illustrated in Figure~\ref{fig:solidangle}, where an annular element of the incoming flux at impact parameter $b$, with thickness $db$, is scattered into an axially symmetric solid angle element associated with the polar angle $\theta$.

The area of this annular region is given by $d\sigma = 2\pi b\, db$. Since $\theta(E,b)$ is an invertible function for fixed $E$, and $\theta$ uniquely determines the solid angle element $d\Omega$ with azimuthal symmetry, one can express the differential cross section using the chain rule as:
\begin{equation}
d\sigma = 2\pi b\frac{db}{d\Omega} \, d\Omega,
\label{dsigma_chainrule}
\end{equation}
where $b=b(\theta)$ implicitly.
In systems with azimuthal symmetry, the element of solid angle can be expressed in terms of the scattering angle as:
\begin{equation}
d\Omega = 2\pi \sin\theta\, d\theta.
\label{domega}
\end{equation}

Substituting \eqref{domega} into \eqref{dsigma_chainrule}, and rearranging, we obtain the differential cross section per unit solid angle:
\begin{equation}
\frac{d\sigma}{d\Omega} = \sum_i \frac{b_i}{\sin\theta} \left|\frac{db}{d\theta}\right|_{b_i} ,
\label{dsigma_domega_sum}
\end{equation}
where the sum runs over all impact parameters $b_i$ that lead to the same scattering angle $\theta$, reflecting the non-monotonicity of the scattering function in attractive potentials.

Although the true differential element is $d\sigma$, we follow standard terminology and refer to $d\sigma/d\Omega$ as the differential cross section. It expresses the angular distribution of scattered particles per unit solid angle and is proportional to the probability density for scattering into direction $\theta$.

For fixed $E$, regions where the function $\theta(E,b)$ varies slowly with $b$ - that is, where $|db/d\theta|$ is large - correspond to angular accumulation of scattered particles. These accumulations can produce sharp peaks in the differential cross section and are associated with the formation of caustics—regions where the density of scattered trajectories diverges geometrically, leading to enhanced intensity in the angular distribution.   

\section{The Lennard-Jones Potential and the Effective Potential}

\begin{figure}[h!]
\centering
\includegraphics[width=0.7\textwidth]{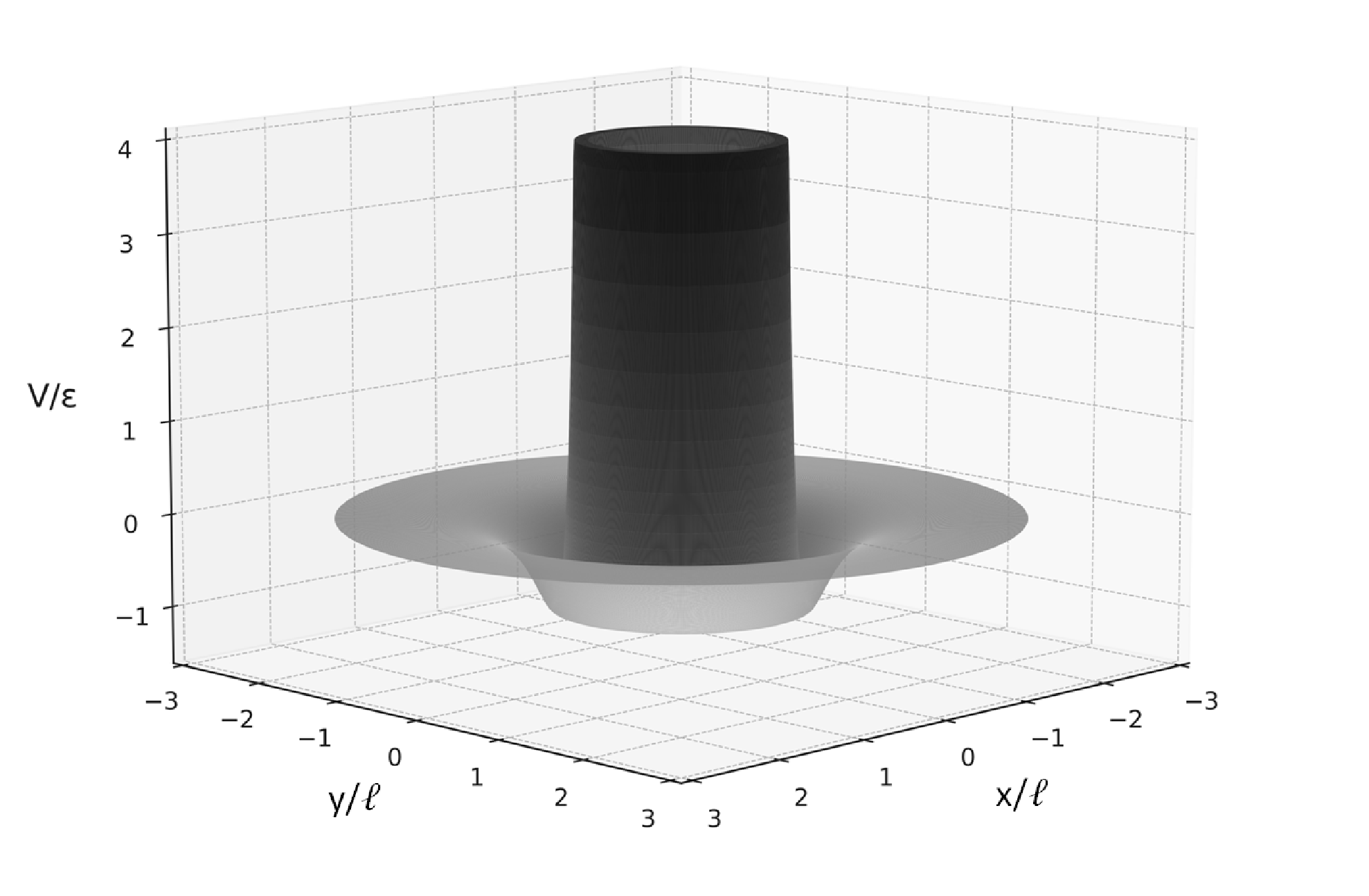}
\caption{Three-dimensional plot of the Lennard-Jones potential in the \(x\)-\(y\) plane, illustrating the steep repulsive core and the broad attractive well.}
\label{fig:pot3D}
\end{figure}

The interaction between neutral, nonpolar atoms or molecules can often be modeled by the Lennard-Jones potential, where the repulsive term $\propto r^{-12}$  represents the short-range Pauli exclusion effects, and the attractive term $\propto r^{-6}$ corresponds to van der Waals forces. A three-dimensional visualization of this potential is shown in Figure~\ref{fig:pot3D}. The surface exhibits a pronounced attractive well, reaching its minimum at a radial distance $r/\ell = 2^{1/6}$, which corresponds to the equilibrium separation between two non-bonded particles. For distances shorter than this, the potential increases sharply due to the strong repulsive interaction, effectively preventing overlap between particles. At large separations, the potential tends asymptotically to zero, reflecting the vanishing influence of intermolecular forces.

In this work, we adopt Lennard-Jones reduced units, setting $\ell = 1$, $\mu=1$ (where $\mu$ denotes the reduced mass allowing the two-body system to be described as an equivalent one-body problem in the relative frame), and expressing energies in units of $\varepsilon$. This simplifies the equations and reduces the number of parameters involved. In these units, time is measured in units of $\ell \sqrt{\mu/\varepsilon}$, distances are measured in units of $\ell$, masses in units of the particle mass. 

To analyze the radial dynamics under a central force, we introduce the effective potential, which allows us to treat the motion as a one-dimensional problem in the radial coordinate. This simplification is possible because of the conservation of angular momentum in a central force field. In this work, we adopt the laboratory frame, where the target is fixed and the projectile moves under the influence of the potential. When the equations of motion are expressed in polar coordinates $(r, \theta)$, the kinetic energy separates into two terms: one associated with radial motion and the other with angular motion. The latter gives rise to a repulsive centrifugal contribution into an effective potential. The resulting expression is:
\begin{equation}
V_{\text{eff}}(r) = V(r) + \frac{L^2}{2r^2},
\label{eq:veff}
\end{equation}
where $V(r)$ is the interaction potential (in our case, the Lennard-Jones potential) and $L$ is the conserved angular momentum of the particle. The term $L^2/2r^2$ is known as the ``centrifugal potential''. Physically, it reflects the tendency of a rotating body to resist inward acceleration.

The key insight is that the dynamics of the particle in $r$ behaves as if it were moving under the influence of a potential \( V_{\text{eff}}(r) \). This framework allows for a geometric interpretation of trajectories and makes it possible to analyze features such as unstable circular orbits, turning points, and orbital barriers using simple energy arguments~\cite{goldstein2002}.

In Cartesian coordinates, the angular momentum is given by $L = x p_y - y p_x$, and since it is conserved and $p_{y_0}=0$, we can write $L^2=L_0^2=2Eb^2$, where $b$ is the impact parameter and $E$ is the total energy of the particle. Thus, the effective potential becomes:
\begin{equation}
V_{\text{eff}}(r) = V(r) + \frac{E b^2}{r^2}.
\label{Veff2}
\end{equation}

\begin{figure}[h!]
\centering
\includegraphics[width=0.7\textwidth]{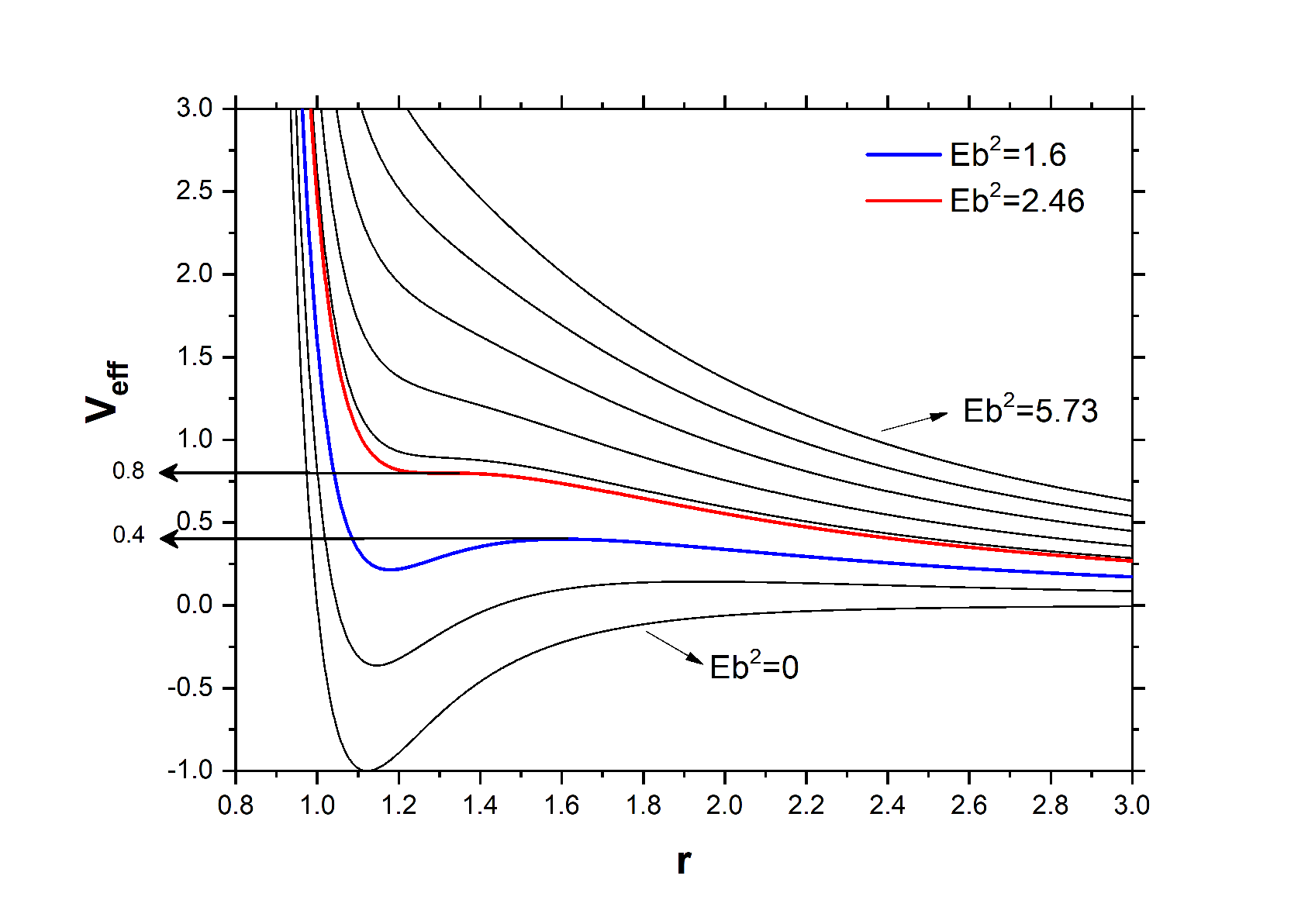}
\caption{Effective potential curves for different values of $E b^2$. The blue and red curves represent $E b^2=1.6$ and $E b^2=2.46$, respectively.}
\label{fig:veff_curves}
\end{figure}

Figure~\ref{fig:veff_curves} shows plots of the effective potential for various values of the product $E b^2$. When $E b^2 = 0$, the curve reduces to the Lennard-Jones potential itself. The blue curve, corresponding to $E b^2 = 1.6$, exemplifies the presence of a local maximum — also observed in the neighboring curves — indicating an unstable circular orbit. The red curve, with $Eb^2 = 2.46$, marks the transition between profiles with and without a local maximum. The energies $V_{\text{eff}} = 0.4$ and $V_{\text{eff}} = 0.8$, marked by horizontal lines in the figure, indicate specific values of the effective potential. The lower one corresponds to the local maximum of the blue curve (for $Eb^2=1.6$) while the upper one corresponds to the inflection point of the red curve (for $Eb^2=2.46$).

Let us consider the condition for the existence of an unstable circular orbit, which corresponds to a local maximum of the effective potential. The position of this maximum, denoted by $r^*$, corresponds to the radius of this orbit and satisfies the condition:
\begin{equation}
\left. \frac{dV_{\text{eff}}}{dr} \right|_{r = r^*} = 0,
\label{Veffmin}
\end{equation}
which leads to:
\begin{equation}
    \left. \frac{dV}{dr} \right|_{r = r^*} = \frac{2Eb_c^2}{{r^*}^3},
\end{equation}
where $b_c$ is the critical impact parameter associated with the unstable circular orbit at $r^*$.

Replacing $dV/dr$ and clearing denominators, we find:
\begin{equation}
2E b_c^2 = \frac{-48 + 24 {r^*}^6}{{r^*}^{10}}.
\label{1}
\end{equation}
From the condition that the total energy equals the effective potential at the maximum, we write:
\begin{equation}
E = V(r^*) + \frac{E b_c^2}{{r^*}^2},
\end{equation}
which leads to:
\begin{equation}
b_c^2 = \frac{E {r^*}^{12} - 4 (1 - {r^*}^6)}{E {r^*}^{10}}.
\label{2}
\end{equation}

Substituting \ref{2} into \ref{1}, we obtain an expression of $r^*$ in terms of $E$:
\begin{equation}
r^* = \left( \frac{4 + 2\sqrt{4 - 5E}}{E} \right)^{1/6}.
\label{eq:rstar_of_E}
\end{equation}

This expression allows us to express $b_c(E)$ by substituting $r^*$ into Equation \ref{2}. It also reveals that for $E > 0.8$, the term under the square root becomes negative, meaning that no real value of $r^*$ exists, and thus the effective potential cannot develop a local maximum. Therefore, unstable circular orbits do not exist for energies greater than this energy threshold.

\section{Classical trajectories and deflection function}

Although the scattering angle  $\theta $ can be expressed in terms of integrals over the particle trajectory, these expressions must still be evaluated numerically for most interaction potentials, including the Lennard-Jones form~\cite{khrapak2014b}, which lacks elementary antiderivatives. For this reason, we opted for a direct numerical integration of the equations of motion, which provides both trajectory data and scattering angles without introducing additional approximations. This approach is particularly valuable when the deflection function becomes non-monotonic, as in the case of orbiting and rainbow scattering phenomena. The system was integrated using a fixed-step fourth-order Runge–Kutta method with a time step $\Delta t = 10^{-4}$. As the problem involves single-pass scattering events rather than long-term orbit propagation, this method yielded highly accurate and stable results, with energy deviations below $10^{-13}$ in the most demanding cases, such as trajectories with prolonged interaction times.

The equations of motion for a particle of unit reduced mass in two-dimensional space under the Lennard-Jones potential are:
\begin{align}
\frac{dx}{dt} &= p_x, \\
\frac{dy}{dt} &= p_y, \\
\frac{dp_x}{dt} &= -\frac{\partial V}{\partial x} = -\frac{dV}{dr} \cdot \frac{x}{r}, \\
\frac{dp_y}{dt} &= -\frac{\partial V}{\partial y} = -\frac{dV}{dr} \cdot \frac{y}{r},
\end{align}
where $r = \sqrt{x^2 + y^2}$, and the radial derivative of the Lennard-Jones potential is given by:
\begin{equation}
\frac{dV}{dr} = 24 \left( \frac{1}{r^7} -\frac{2}{r^{13}}\right).    
\end{equation}

The initial conditions were chosen such that the particle starts at a horizontal position $x_0 = -10$, sufficiently far from the scatterer so that the potential is negligible, and at a vertical position $y_0 = b$, which defines the impact parameter. The initial momentum is aligned with the horizontal axis, with components ${p_x}_0 = \sqrt{2E}$ and ${p_y}_0 = 0$, corresponding to a total energy $E$.

The integration proceeds until $t_{tot}$, when the particle moves beyond the initial radial distance, that is, when $r(t_{tot}) > r(0)$, indicating that the interaction has effectively ended. At that point, the final scattering direction is determined from the momentum vector, and the scattering angle is computed as
\begin{equation}
\theta = \left| \arctan\left( \frac{p_y(t_{tot})}{p_x(t_{tot})} \right) \right|.
\end{equation}

This formula corresponds to the modulus of the angle formed between the momentum vector and the positive horizontal axis. The computation considers the correct quadrant via the two-argument arctangent function (typically implemented as \texttt{atan2}), which returns values in the interval $[-\pi, \pi]$, allowing the full angular range to be accessed before applying the modulus. We opted for this approach instead of using the expression $\arccos(p_x/\sqrt{2E})$, which is limited to the interval $[0, \pi]$, to ensure access to the geometric scattering angle $\theta_{\text{geom}}$ in $[-\pi, \pi]$ before taking the absolute value.

In order to investigate orbiting and divergent behavior in more detail, we introduce the deflection angle  $\varphi$, defined as:
\begin{equation}
\varphi = \theta_{\text{geom}} - 2n\pi,
\end{equation}
where $n$ counts the number of complete revolutions the particle performs around the target, measured by the number of upward crossings of the $x=0$ axis in the region $x<0$. This criterion allows $\varphi$ to track the effective accumulated angular deviation of the particle due to the interaction, even if it completes multiple revolutions.

The sign and magnitude of $\varphi$ carry essential physical meaning. For $b > 0$, all trajectories approach from above the scatterer, and any rotation of the trajectory occurs in the clockwise direction. Consequently, $\varphi$ takes negative values, and in cases of strong interaction, it can drop below $-\pi$. This continuous definition of $\varphi$ makes it suitable for analyzing orbiting regimes and the structure of the deflection function.

\begin{figure}[h!]
\centering
\includegraphics[width=0.6\textwidth]{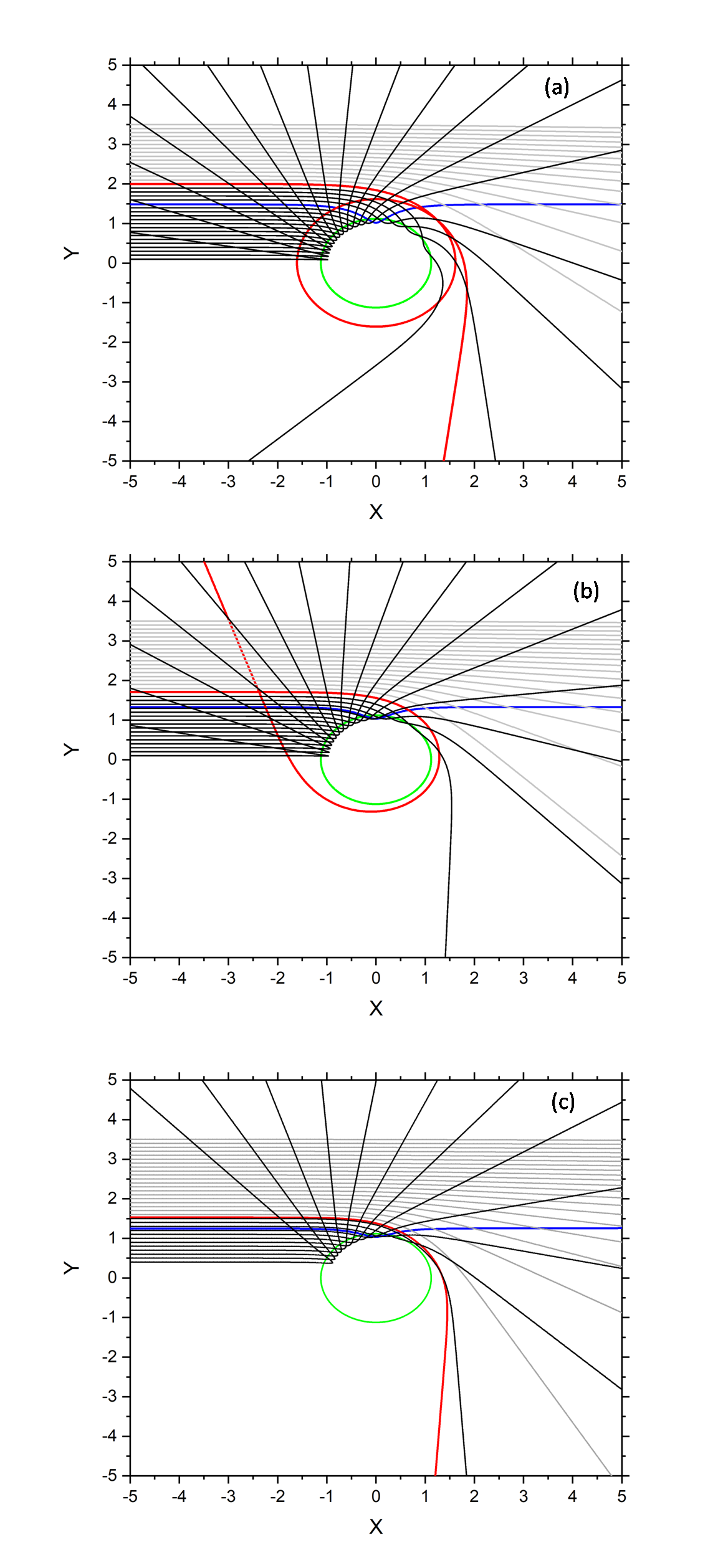}
\caption{Trajectories for $E=0.4$, $E=0.9$, and $E=1.5$, with critical impact parameters $b_c \approx 2.0$, $b_c \approx 1.7$, and $b_r \approx 1.5$, respectively.}
\label{fig:trajet}
\end{figure}

Figure~\ref{fig:trajet} shows representative sets of trajectories for three distinct energies: (a) $E = 0.4$, (b) $E = 0.9$, and (c) $ E= 1.5$. These energies were chosen to represent qualitatively different regimes of the scattering dynamics: below the critical energy $E = 0.8$, just above it, and significantly beyond it. The green circle in each panel corresponds to $r = 2^{1/6}$, which marks the boundary between the attractive and repulsive regions of the Lennard-Jones potential. The red trajectory represents the critical impact parameter: $b = b_c \approx 2.0$ for $E = 0.4$, where the particle asymptotically approaches an unstable circular orbit, and $b = b_r \approx 1.7$ and $b_r \approx 1.5$ for $E = 0.9$ and $E = 1.5$, respectively. The value $b=b_r$ is defined as the impact parameter for which the deflection angle $\varphi(b)$ reaches its minimum, leading to a local accumulation of scattered particles at the corresponding rainbow scattering angle $\theta_r$. Black curves correspond to $b < b_c$ or $b < b_r$, while gray curves denote  $b > b_c$ or $b > b_r$. The separation between black and gray emphasizes the non-monotonicity of the scattering angle $\theta$ as a function of the impact parameter: the same value of $\theta$ may be reached from different values of $b$, indicating the presence of multivalued regions in the deflection function.

\begin{figure}[h!]
\centering
\includegraphics[width=0.7\textwidth]{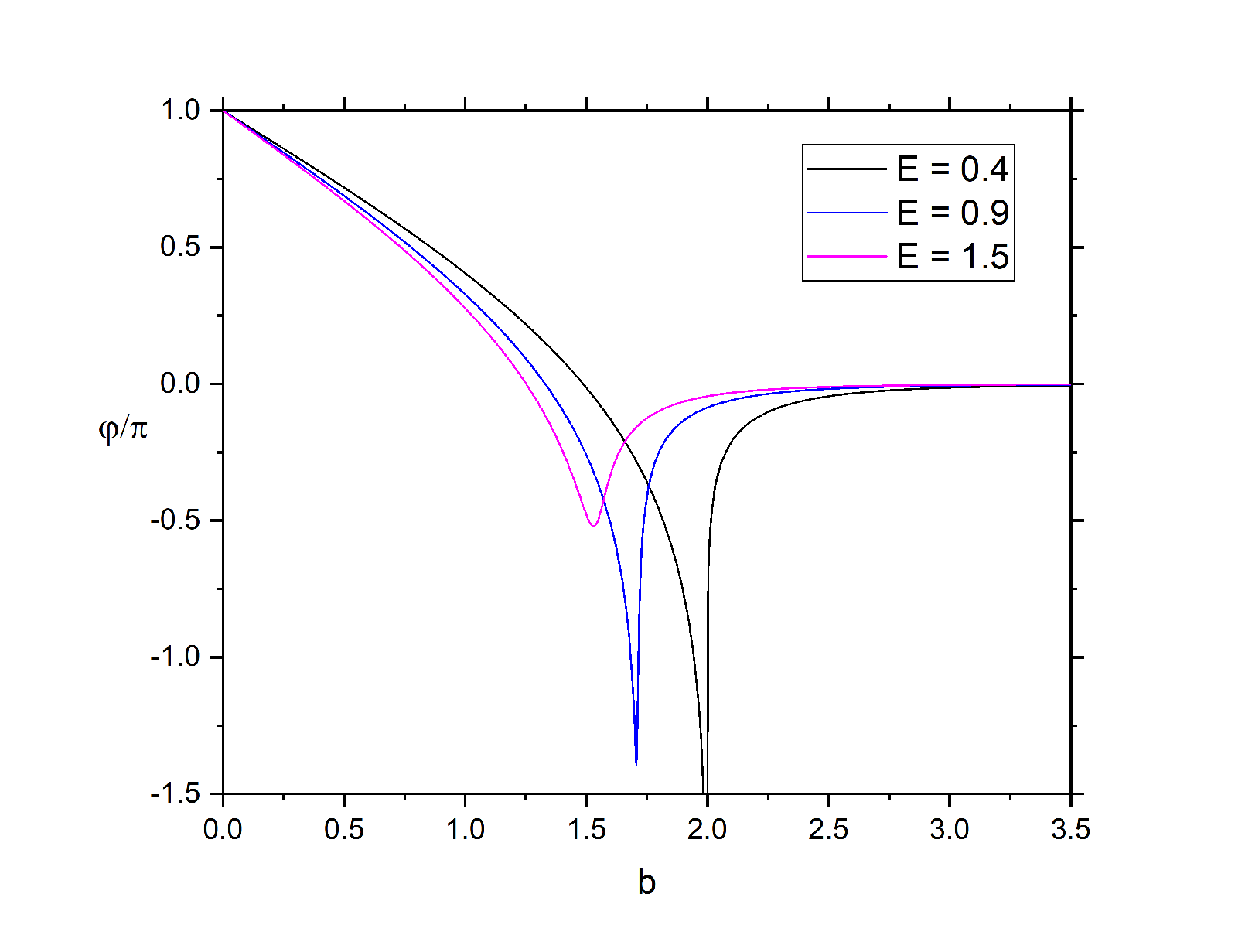}
\caption{Deflection angle $ \varphi(b) $ for $ E = 0.4 $, $ E = 0.9 $, and $ E = 1.5 $.}
\label{fig:deflete}
\end{figure}

The deflection angle as a function of the impact parameter is shown in Figure~\ref{fig:deflete}. Although the scattering angle $ \theta $ is the one used in the calculation of the cross section in Equation ~\eqref{dsigma_domega_sum}, its inverse relation $ b(\theta) $ shares the same absolute derivative as the deflection curve $ \varphi(b) $, that is, $ |db/d\theta| = |db/d\varphi| $, since both angles are related by a continuous and monotonic transformation. Therefore, the features observed in the deflection plots—such as minima and divergences—correspond directly to the features relevant for the evaluation of $d\sigma/d\Omega$, justifying the use of $ \varphi(b) $ curves for visual reference.

Each curve corresponds to the same energies as in the previous figure. For $E = 0.4$, the deflection angle diverges negatively as $b \to b_c \approx 2.0$, which corresponds to the onset of an unstable orbit (red curve in Figure~\ref{fig:trajet}a). For $E = 0.9$ and $E = 1.5$, the deflection angle remains finite but exhibits a local minimum at $b = b_r \approx 1.7$ and $b_r \approx 1.5$, respectively (red curves in Figures~\ref{fig:trajet}b and~\ref{fig:trajet}c). These minima correspond to rainbow-type scattering and cause the differential cross section $d\sigma/d\Omega$, defined in Equation~\eqref{dsigma_domega_sum}, to diverge due to the vanishing of the derivative $d\theta/db$.

The blue trajectories shown in Figure~\ref{fig:trajet} correspond to the first crossing of the deflection functions $\varphi(b)$ through zero, which translates into $\theta = 0$ in the scattering angle. This crossing occurs smoothly and thus leads to a divergence of $d\sigma/d\Omega$ at $\theta = 0$. This phenomenon is known as forward glory scattering and is always present due to the asymptotic behavior $\varphi(b) \to 0$ as $b \to \infty$.

A similar divergence may occur at $\theta = \pi$, referred to as backward glory scattering. It is important to emphasize that this phenomenon is not associated with the trivial backscattering produced at $b = 0$, where the particle is reflected directly backward. Although such head-on collisions do result in $\theta = \pi$, they do not produce a divergence in $d\sigma/d\Omega$ because the differential cross section is proportional to $b$, which vanishes at the origin. In contrast, backward glory scattering arises from trajectories with nonzero impact parameter that encircle the target, such that the deflection angle satisfies $\varphi = -\pi$, eventually producing $\theta = \pi$ through the transformation rule. In the present data, this condition is fulfilled for $E = 0.4$ and $E = 0.9$, but not for $E = 1.5$, as can be seen from the shape of the deflection curves.

\section{Scattering angle, delay, and differential cross section analysis}

In the previous section, we analyzed classical trajectories and the deflection angle $\varphi$ as a function of the impact parameter $b$. We observed that for energies below the critical value $E = 0.8$, the deflection diverges at a critical impact parameter $b_c$, associated with an unstable circular orbit. For $E > 0.8$, there is no longer a divergence, but a minimum appears in the deflection function at $b = b_r$ , indicating the emergence of rainbow-type scattering. 

In this section, we extend the analysis to include three key aspects: the scattering angle $\theta$, the relative interaction delay $\tau$, and the differential cross section $d\sigma/d\Omega$. First, we construct color maps of the scattering angle $\theta$ in the $E - b$ plane to explore how the measurable scattering direction depends on energy and impact parameter. We then analyze the interaction delay by comparing the total trajectory time $t_{\text{tot}}$ with the transit time $t_{\text{tran}}$ of a straight-line motion in the absence of interaction. The relative delay $\tau = (t_{\text{tot}} - t_{\text{tran}})/t_{\text{tran}}$ highlights regions where the particle experiences significant time retardation due to orbiting or near-trapping events. Finally, we compute the differential scattering cross section $d\sigma/d\Omega$, identifying the contributions from multivalued regions of $b$, and interpret the associated singularities in terms of glory and rainbow scattering phenomena.

\begin{figure}[h!]
\centering
\includegraphics[width=0.7\textwidth]{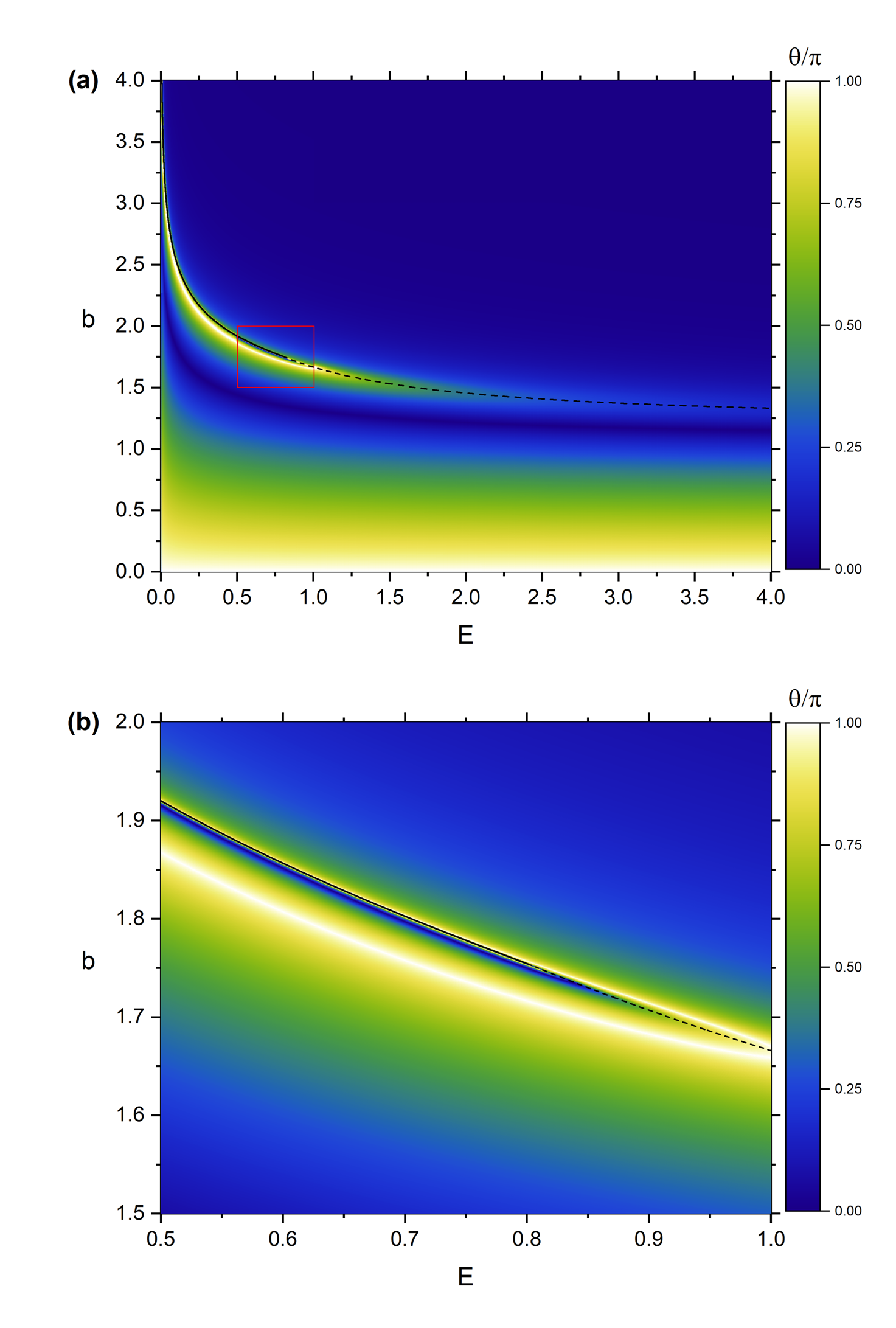}
\caption{(a) Scattering angle $\theta$ in the $E - b$ plane. (b) Magnified view of the region near the critical transition.}
\label{fig:esc}
\end{figure}

Figure~\ref{fig:esc}(a) presents a color map of the scattering angle $ \theta $ as a function of energy $ E $ and impact parameter $ b $. The color scale ranges from white (angles close to $ \pi $) to dark blue (angles close to zero), reflecting the degree of deflection experienced by the particle. For large $ b $, the angle tends to zero due to negligible interaction. As $ b $ decreases, the deflection grows, eventually reaching $ \theta = \pi $ for head-on collisions. The solid curve traces the values of $ b $ that produce a divergence in the deflection angle, corresponding to the critical line $ b_c(E) $ for $ E < 0.8 $ given by Equation \ref{2}, while the dashed curve follows the values of $ b_r(E) $ for $ E > 0.8 $, associated with the minima in the deflection function.

The absence of an analytical expression for $b_r$ reflects its dynamic origin: while $b_c$ arises from a singularity in the effective potential, the values of $b_r$ must be determined numerically as the points of maximum (in absolute value) deflection. The transition between the two regimes occurs at $E = 0.8$, where the topology of the deflection function changes. This is further clarified in the magnified view of the critical region shown in Figure~\ref{fig:esc}(b), where the continuity between the $b_c$ and $b_r$ curves becomes apparent. There, the curves $b_c(E)$ and $b_r(E)$ clearly delineate the boundary between two distinct branches of $b$ that correspond to the same scattering angle.

\begin{figure}[h!]
\centering
\includegraphics[width=0.7\textwidth]{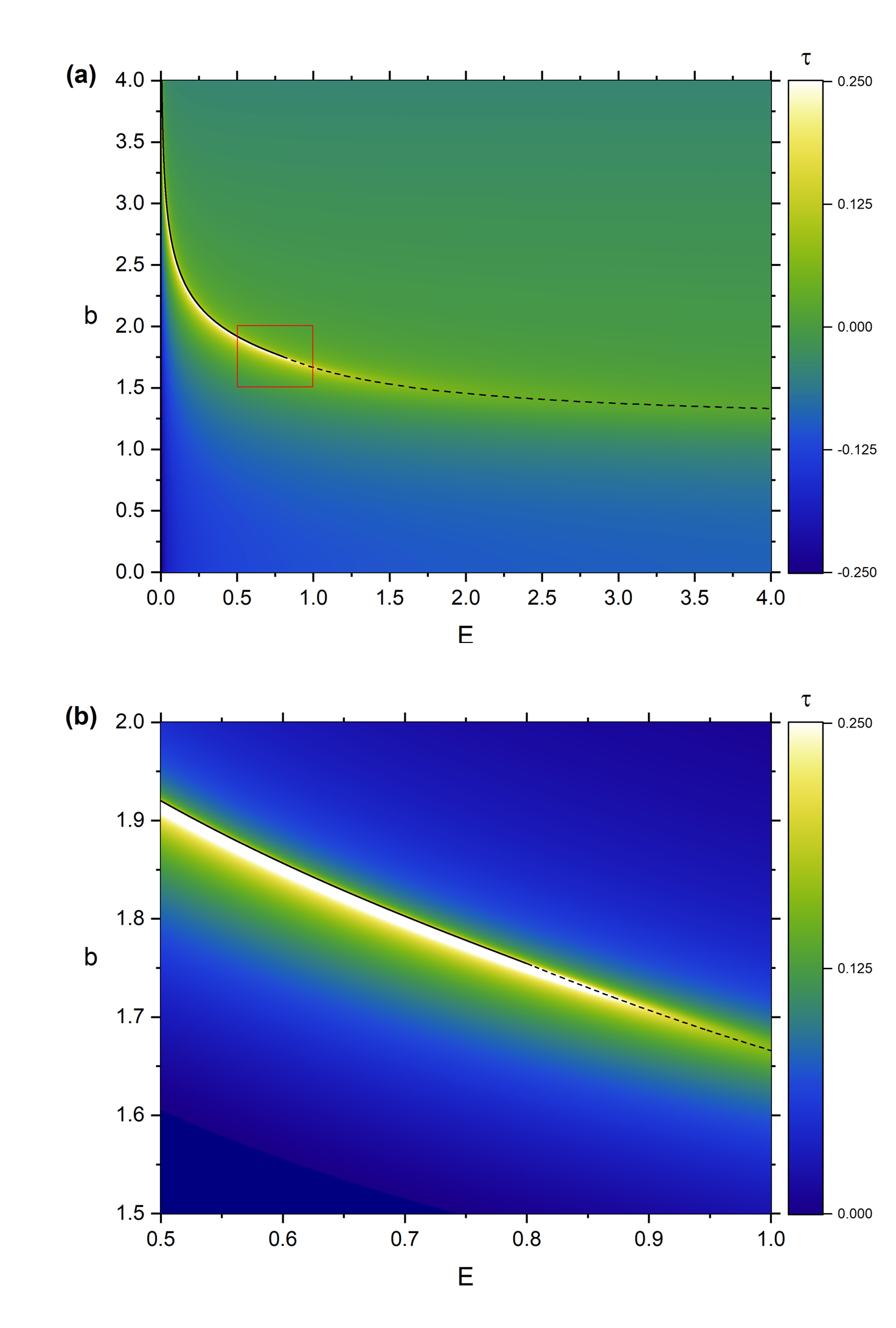}
\caption{(a) Relative delay $ \tau $ in the $ E - b $ plane. (b) Magnified view with scale restricted to $ 0 \leq \tau \leq 0.25 $.}
\label{fig:delay}
\end{figure}

The dynamic structure revealed by the scattering angle map is mirrored in the interaction delay. Figure~\ref{fig:delay}(a) shows the relative delay $\tau$. The color scale ranges from $-0.25$ (dark blue) to $0.25$ (white), while values below $-0.25$ and above $0.25$ are saturated at the endpoints (dark blue and white, respectively). Positive values indicate that the scattered particle takes longer than the transit time, while negative values represent faster passage. The curve $b_c(E)$ for $E < 0.8$ and its continuation $b_r(E)$ for $E > 0.8$ appear as a sharp ridge in the delay landscape, highlighting that these conditions correspond to orbiting behavior and prolonged interaction.

A detailed view of the transition zone is shown in Figure~\ref{fig:delay}(b), using a restricted scale from $0$ to $0.25$. Values exceeding this range are plotted in white, while negative delays are rendered in dark blue. The fine agreement between the features in Figures~\ref{fig:esc}(b) and~\ref{fig:delay}(b) confirms that regions of extreme angular behavior are accompanied by enhanced temporal trapping, offering a complementary perspective on the dynamics of scattering in central potentials.

The differential scattering cross section $d\sigma/d\Omega$ was obtained from the definition given in Equation~\eqref{dsigma_domega_sum}. However, its numerical implementation required special care. Since the derivative $ \left| \frac{db}{d\theta} \right| $ had to be evaluated from the deflection curves $ \varphi(b) $, and we have direct control over the sampling in $ b $, but not in $ \theta $, the distribution of calculated points is inherently nonuniform in $ \theta $. To overcome this, we first computed the deflection function using a dense and finely spaced grid of $ b $-values. Then, we interpolated the resulting points to construct an approximately uniform and sufficiently refined sampling in $ \theta $, from which we could estimate the derivative $ db/d\theta $ using finite differences.

Moreover, in regions where the function $ \theta(b) $ is non-monotonic, multiple values of $ b $ correspond to the same scattering angle $ \theta $. In such cases, the total cross section must include the sum of all contributions, and this was accounted for by segmenting the function $ \theta(b) $ into monotonic branches prior to computing $ \left| \frac{db}{d\theta} \right| $. This ensured that every rainbow-related divergence was properly captured and that the multivalued structure of the scattering process was faithfully represented.

\begin{figure}[htb]
\centering
\includegraphics[width=0.75\columnwidth]{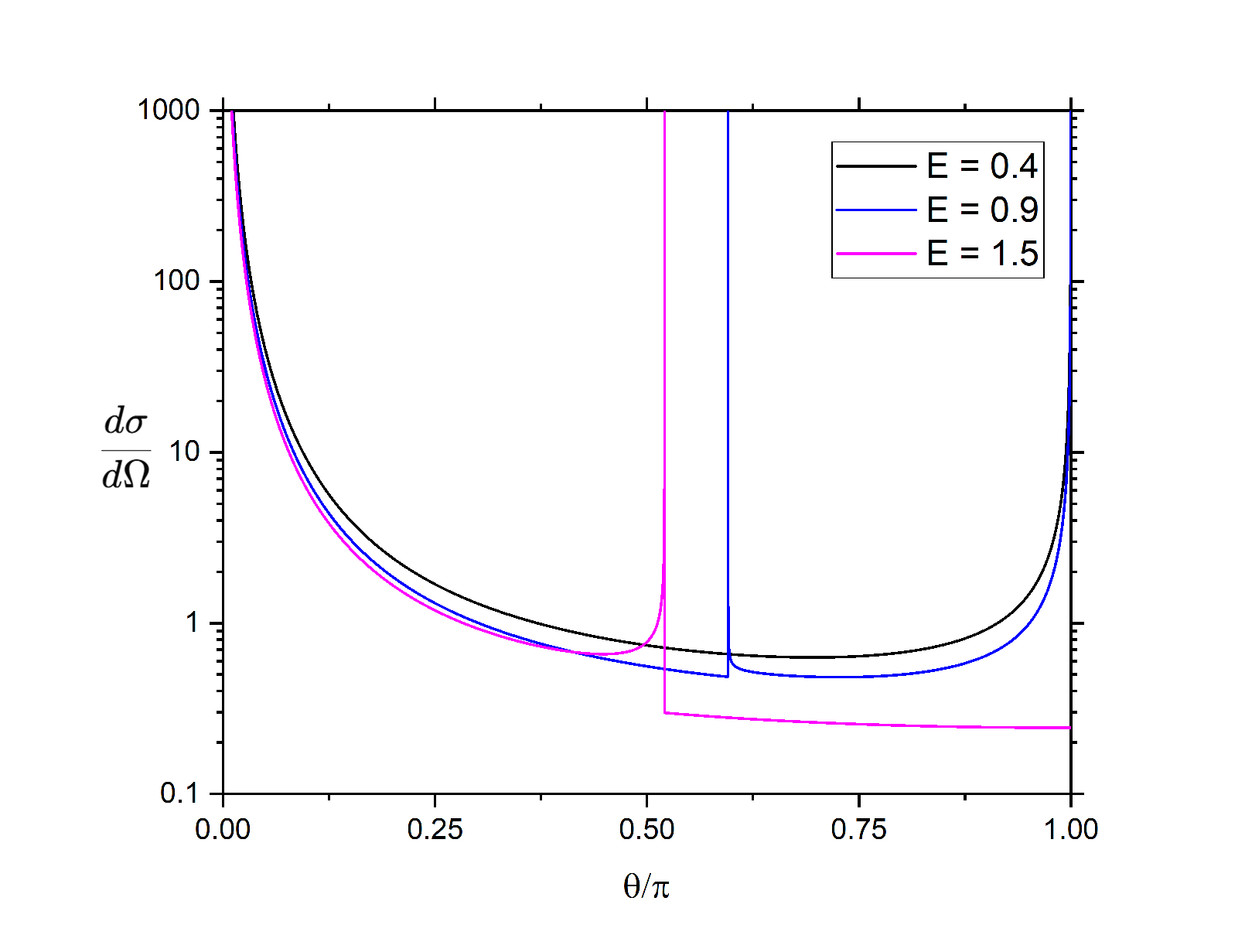}
\caption{Differential cross section $ \sigma(\theta) $, in logarithmic scale, for three energy values: $ E = 0.4 $ (black), $ E = 0.9 $ (blue), and $ E = 1.5 $ (magenta).}
\label{fig:cross}
\end{figure}

Figure~\ref{fig:cross} shows the cross section curves $d\sigma/d\Omega$ as a function of scattering angle for three different energies. All curves exhibit a steep rise in the vicinity of $ \theta = 0 $, characterizing the forward glory scattering, associated with trajectories suffering minimal deflection and concentrating near the forward direction.
A second divergence appears at $ \theta = \pi $ for the curves with $ E = 0.4 $ and $ E = 0.9 $, but is absent for $ E = 1.5 $. This effect, known as backward glory scattering, occurs when the deflection angle $ \varphi(b) $ reaches or crosses $ -\pi $, which does not occur for $ E = 1.5 $ as can be seen in Figure~\ref{fig:deflete} and will only happen for $ E \lesssim 1 $ as will be verified. The resulting divergence in $ d\sigma/d\Omega $ is due to the vanishing sine term in the denominator of the cross section formula.

For the $ E = 0.9 $ and $ E = 1.5 $ cases, a third divergence appears at intermediate angles $ \theta_r \approx 0.65\pi $ and $ \theta_r \approx 0.55\pi $, respectively. These peaks are signatures of the rainbow scattering phenomenon, caused by the non-monotonicity of the deflection function $ \varphi(b) $ near the critical value $ b = b_r $, where $ \varphi(b_r) = \varphi_{\text{min}} $ and consequently $ d\varphi/db = 0$, leading to $d\theta/db = 0$ and causing a divergence in Equation~(\ref{dsigma_domega_sum}). 
Finally, for $ E = 1.5 $, the curve presents a relatively flat and low plateau for $ \theta > \theta_r $, indicating that beyond the rainbow angle, the distribution of scattered particles becomes smoother and more uniform.

\begin{figure}[htb]
\centering
\includegraphics[width=0.75\columnwidth]{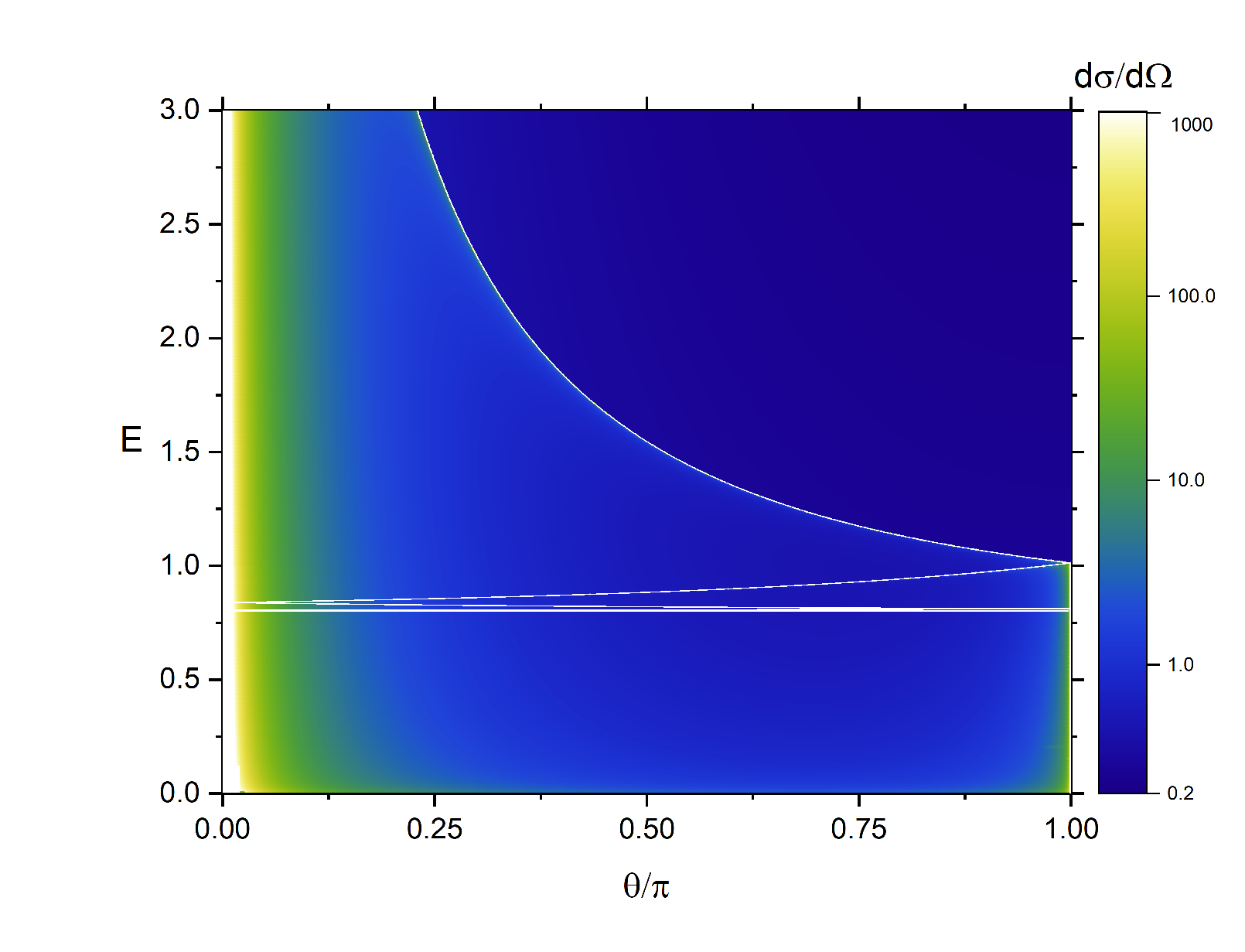}
\caption{Color map of the differential cross section $d\sigma/d\Omega$ in the $\theta-E$ plane.}
\label{fig:parcross}
\end{figure}

The map shown in Figure~\ref{fig:parcross} offers a broad visualization of the differential cross section across a range of energies and scattering angles. The color scale is logarithmic, ranging from $0.2$ (dark blue) to $1000$ (white), with values below and above these thresholds saturated at the respective endpoints to enhance contrast across several orders of magnitude. The figure shows vertical bands of high differential cross section near the limits $ \theta = 0 $ and $ \theta = \pi $. The bright stripe near $ \theta = 0 $ corresponds to forward glory scattering, a robust feature present for all energies, resulting from particles being weakly deflected and concentrating in the forward direction. Near $ \theta = \pi $, a much thinner and energy-restricted band can also be observed for $ E \lesssim 1 $, indicating the presence of backward glory scattering. Due to the narrowness of this effect, the corresponding region appears less intense in the color scale, though it still reflects a strong enhancement of $ d\sigma/d\Omega $ at $ \theta \approx \pi $. The high values of $ d\sigma/d\Omega $ associated with rainbow scattering appear as a white ridge, which gradually flattens as the energy approaches $ E = 0.8 $, accumulating near this critical value. This horizontal accumulation reflects the emergence of quasi-orbital trajectories with extremely large deflections.

Just before the ridge, a band of light blue indicates a rapid increase in $ d\sigma/d\Omega $ as $ \theta $ approaches $ \theta_{\text{r}} $. In contrast, the dark blue region immediately after the ridge corresponds to a sharp drop in the scattering cross section immediately beyond $ \theta_r $, where the angular density of scattered particles rapidly decreases. The transformation from the deflection angle $ \varphi $ to the laboratory scattering angle $ \theta $ (constrained to $ \theta \in [0, \pi] $) leads to folding and accumulation effects. This transformation gives rise to a wave-like pattern in the range $ 0.8 < E \lesssim 1 $, where $ \theta_r $ exhibits a large rate of change within the interval $ [0, \pi] $, since the most deflected trajectory for each $ E $ increasingly resembles a circular orbit as $ E \to 0.8 $. This behavior is ultimately a manifestation of the topological change in the scattering regime as the energy crosses the threshold at $ E = 0.8 $.

\section{Conclusions}

This study revisits the classical scattering problem under a Lennard-Jones potential with an emphasis on conceptual clarity and pedagogical relevance. Through direct numerical integration of the equations of motion, we explored the behavior of deflection angles, particle trajectories, interaction delays, and differential cross sections across a range of energies and impact parameters.

We identified two qualitatively distinct scattering regimes: divergence of the deflection angle for $E < 0.8$, associated with unstable circular orbits, and rainbow scattering for $E > 0.8$, linked to a minimum in the deflection function. These phenomena were illustrated using color maps and numerical data, enabling visual interpretations of key concepts such as orbiting, delay, and multivalued angular distributions.

By combining classical mechanics, effective potential analysis, and modern visualization tools, this work provides an accessible framework for understanding central force scattering. The results may serve as a valuable resource for students and instructors seeking to connect theoretical models with computational and graphical insights.


\begin{thebibliography}{99}
\bibitem{zugec2020} P. Žugec and I. Topić, \textit{A shadow of the repulsive Rutherford scattering}, Eur. J. Phys. \textbf{41}, 065005 (2020).
\bibitem{lennardjones1924} J. E. Lennard-Jones, \textit{On the determination of molecular fields. I. From the variation of the viscosity of a gas with temperature}, Proc. R. Soc. Lond. A \textbf{106}, 441 (1924).
\bibitem{hirschfelder1948} J. O. Hirschfelder, R. B. Bird, and E. L. Spotz, \textit{The transport properties for non-polar gases}, J. Chem. Phys. \textbf{16}, 968 (1948).
\bibitem{stephan2019} S. Stephan, M. Thol, J. Vrabec, and H. Hasse, \textit{Thermophysical Properties of the Lennard-Jones Fluid}, J. Chem. Inf. Model. (2019).
\bibitem{nietiadi2022} M. L. Nietiadi and H. M. Urbassek, \textit{Bouncing and spinning of amorphous Lennard-Jones nanoparticles under oblique collisions}, Sci. Rep. \textbf{12}, 10699 (2022).
\bibitem{sperandeo1987} R. M. Sperandeo-Mineo and G. Tripi, \textit{Microcomputer simulation of a two-dimensional Lennard-Jones fluid}, Eur. J. Phys. \textbf{8}, 117–124 (1987).
\bibitem{venkattraman2012} A. Venkattraman and A. A. Alexeenko, \textit{Binary scattering model for Lennard-Jones potential: Transport coefficients and collision integrals for non-equilibrium gas flow simulations}, Phys. Fluids \textbf{24}, 027101 (2012).
\bibitem{khrapak2014b} S. A. Khrapak, \textit{Accurate transport cross sections for the Lennard-Jones potential}, Eur. Phys. J. D \textbf{68}, 276 (2014).
\bibitem{laricchiuta2007} A. Laricchiuta et al., \textit{Classical transport collision integrals for a Lennard-Jones like phenomenological model potential}, Chem. Phys. Lett. \textbf{445}, 133–139 (2007).
\bibitem{pirani2008} F. Pirani et al., \textit{Beyond the Lennard-Jones model}, Phys. Chem. Chem. Phys. \textbf{10}, 5489–5503 (2008).
\bibitem{khrapak2014c} S. A. Khrapak, \textit{Classical scattering in strongly attractive potentials}, Phys. Rev. E \textbf{89}, 032145 (2014).
\bibitem{goldstein2002} H. Goldstein, C. Poole, and J. Safko, \textit{Classical Mechanics}, 3rd ed., Addison–Wesley (2002).
\bibitem{balint2015} G. G. Balint-Kurti and A. Palov, \textit{Theory of Molecular Collisions}, Royal Society of Chemistry (2015).
\bibitem{Landau} L. D. Landau and E. M. Lifshitz, \textit{Mechanics}, , 3rd ed., Butterworth-Heinemann (1976).
\end{thebibliography}
\end{document}